\documentclass[12pt]{article}
\usepackage{amssymb}
\usepackage{amsmath}
\usepackage{amstext}
\usepackage{graphicx,epsfig}
\usepackage{epsfig}
\usepackage{verbatim} 
\usepackage{fancyhdr}
\usepackage{fancybox}
\usepackage{color}
\usepackage{ulem,bbold}
\usepackage{enumitem}
\usepackage{subfigure}
\usepackage{bbm}
\usepackage{parskip}
\usepackage{cite}

\newcommand{\Comment}[1]{{}}
\definecolor{MyDarkBlue}{rgb}{0.15,0.15,0.45}
\usepackage[linktocpage=true]{hyperref}
\hypersetup{
colorlinks=true,
citecolor=MyDarkBlue,
linkcolor=MyDarkBlue,
urlcolor=MyDarkBlue,
}

\newcommand{\be}{\begin{equation}}
\newcommand{\ee}{\end{equation}}
\newcommand{\bea}{\begin{eqnarray}}
\newcommand{\eea}{\end{eqnarray}}
\newcommand{\beas}{\begin{eqnarray*}}
\newcommand{\eeas}{\end{eqnarray*}}
\newcommand{\nn}{\nonumber}

\numberwithin{equation}{section}

\begin{document}


\begin{center}
{\Large \bf{A Note on the AdS/CFT Equivalence Transformation \\}}
\vspace{.2cm}
{\Large \bf{for Galileons in General Dimensions}}
\end{center} 
 \vspace{1truecm}
\thispagestyle{empty} \centerline{
{\large Kurt Hinterbichler}$^{}$\footnote{E-mail: \Comment{\href{mailto:kurt.hinterbichler@case.edu}}{\tt kurt.hinterbichler@case.edu}},\ \ {\large Samanta Saha}$^{}$\footnote{E-mail: \Comment{\href{mailto:sxs2638@case.edu}}{\tt sxs2638@case.edu}} 
}

\vspace{1cm}

\centerline{{\it ${}^{}$CERCA, Department of Physics,}}
\centerline{{\it Case Western Reserve University, 10900 Euclid Ave, Cleveland, OH 44106}}

 \vspace{1cm}

\begin{abstract}

The AdS/CFT equivalence transformation is a field redefinition that relates the Weyl dilaton and AdS brane realizations of broken conformal symmetry.  Acting on theories with second order equations of motion, it maps the conformal galileons to the DBI galileons for a flat brane probing an AdS bulk.  Here we extend this map to arbitrary dimensions and resolve an apparent puzzle regarding the Wess-Zumino terms in odd dimensions.

\end{abstract}

\newpage

\thispagestyle{empty}

\setcounter{page}{1}
\setcounter{footnote}{0}

\parskip=5pt
\normalsize

\section{Introduction}

When conformal symmetry is spontaneously broken to Poincar\'e symmetry, the resulting low energy effective field theory (EFT) contains a scalar Goldstone mode known as the dilaton.  There are an infinite number of ways in which to parametrize the dilaton EFT, but there are two common parameterizations that are particularly useful and well adapted to common situations.

The first is the Weyl realization.  The EFT in this realization is built by constructing curvature invariants made from the conformally flat metric 
\be g_{\mu\nu}=e^{2\pi}\eta_{\mu\nu}\, ,\label{confflatdme}\ee
where $\pi$ is the dilaton field.  The only relevant curvature tensor is the Ricci scalar 
\be R_{\mu\nu}(g) = (D-2)\left( \partial_\mu\pi\partial_\nu\pi-\partial_\mu\partial_\nu\pi-(\partial\pi)^2\eta_{\mu\nu}\right)-\square\pi\eta_{\mu\nu}\,,\label{riccite}\ee
because the Riemann tensor can be expressed in terms of the Ricci tensor on the conformally flat metric \eqref{confflatdme}.  The most general invariant Lagrangian is then a scalar function ${\cal F}$ made from these ingredients, as well as the covariant derivative $\nabla_\mu$ associated to \eqref{confflatdme},
\be {\cal L}=\sqrt{-g}{\cal F}\left(g_{\mu\nu},\nabla_\mu,R_{\mu\nu}\right).\label{dilatongenformee}\ee
As we review below, there can also be a term, the Wess-Zumino term, that cannot be written in terms of the invariant building blocks as in \eqref{dilatongenformee}, even up to a total derivative, but is nevertheless still invariant up to a total derivative under the broken conformal symmetries.

The second is the anti-de Sitter (AdS) realization.  The EFT in this realization is constructed from a flat $D$ dimensional brane probing a fixed $D+1$ dimensional AdS geometry of radius $L$.  We parametrize the brane geometry in terms of a brane bending mode $\phi$ that measures transverse fluctuations away from the flat configuration.  The geometric ingredients are the induced metric on the brane,
\be   \hat g_{\mu\nu}= e^{2\phi}\eta_{\mu\nu}+L^2 \partial_\mu\phi\partial_\nu\phi \,,\label{adsmetrfee}\ee
and the extrinsic curvature of the brane,
\be K_{\mu\nu} ={1\over L\sqrt{1+L^2 e^{-2\phi}(\partial\phi)^2}}\left(e^{2\phi}\eta_{\mu\nu}-L^2\partial_\mu\partial_\nu\phi+2L^2\partial_\mu\phi\partial_\nu\phi\right) \,. \label{braneexcuree}\ee
The intrinsic curvature invariants made from the Riemann tensor of \eqref{adsmetrfee} can be ignored because they can be expressed in terms of the extrinsic curvature via the Gauss–Codazzi relations,
\be \hat R_{\mu\nu\rho\sigma}=K_{\mu\rho}K_{\nu\sigma}- K_{\mu\sigma}K_{\nu\rho} -{1\over L^2}\left(  \hat g_{\mu\rho}\hat g_{\nu\sigma}- \hat g_{\mu\sigma}\hat g_{\nu\rho}  \right) \,.\label{gaussdozadse2}\ee
The most general invariant Lagrangian is then a scalar function ${\cal F}$ made from these ingredients, including the covariant derivative $\hat \nabla_\mu$ associated to \eqref{adsmetrfee},
\be {\cal L}=\sqrt{-\hat g}{\cal F}\left(\hat g_{\mu\nu},\hat \nabla_\mu,K_{\mu\nu}\right).\label{dbigbekrlee}\ee
As is the dilaton realization, there can be a term, the Wess-Zumino term, that cannot be written in terms of the invariant building blocks as in \eqref{dbigbekrlee} up to a total derivative, but is still invariant up to a total derivative under the broken conformal symmetries.

More details of these two realizations, including how they arise from the coset construction and how the broken conformal symmetry acts, can be found in \cite{Hinterbichler:2023nyz}, whose conventions we follow here, as well as \cite{Goon:2011qf,Goon:2011uw,Trodden:2011xh,Goon:2012dy}

In \cite{Bellucci:2002ji}, an explicit field re-definition was found that connects these two realizations, demonstrating that they are different parameterizations of the same EFT.  They called this map between the two realizations the ``AdS/CFT equivalence transformation.''  The map stems from using different parametrizations in the coset construction of the EFT.  It involves infinite powers of the fields and derivatives and is thus essentially non-local, but it is invertible and is local order-by-order in an expansion in powers of the field, so it preserves the S-matrix and does not spoil any of the locality properties of the EFT.  The AdS/CFT equivalence transformation shows that the parameter $L^2$ in the AdS parametrization is redundant (in fact, by a similar transformation $L^2$ can even be made negative, leading to an equivalent theory of a flat brane probing a de Sitter space with two time dimensions \cite{Hinterbichler:2023nyz}).

If we demand that the EFT yield second order equations of motion so that it is free of ghostly degrees of freedom non-linearly \cite{Ostrogradsky:1850fid,deUrries:1998obu}, then it turns out that there are only a finite number of allowed terms, known as galileon terms.   In $D=4$, there are five galileon terms.  In the AdS parametrization their Lagrangians can be written as
\bea   
 {\cal L}_1^{\rm (AdS)} &=&{1\over 4L^4} e^{4\phi} \, ,\nn\\
 {\cal L}_2^{\rm (AdS)} &=&-{1\over L^4}\sqrt{- \hat g} \ ,\nn\\
 {\cal L}_3^{\rm (AdS)} &=&{1\over L^3}  \sqrt{- \hat g}K\,,\nn\\
 {\cal L}_4^{\rm (AdS)} &=&-{1\over L^2} \sqrt{- \hat g} \left[ -  K^2+ K_{\mu\nu}^2 +{12\over L^2} \right]  \ ,\nn\\
 {\cal L}_5^{\rm (AdS)} &=&{1\over L}  \sqrt{- \hat g} \left[ K^3-3K K_{\mu\nu}^2+2K_{\mu\nu}^3 -{9\over L^2}K \right]   \,,\ 
\label{ghostfreegentermsadsp1} 
\eea
where indices are moved with $\hat g_{\mu\nu}$.  All of these terms, except for ${\cal L}_1^{\rm (AdS)}$, are of the form \eqref{dbigbekrlee}, and as shown in~\cite{deRham:2010eu} they correspond to the possible Lovelock terms~\cite{Lovelock:1971yv} on the brane and Gibbons-Hawking-type boundary terms  \cite{Gibbons:1976ue,York:1972sj,Myers:1987yn,Dyer:2008hb} corresponding to the possible Lovelock terms in the bulk.  The non-trivial Lovelock invariants on the four dimensional brane are the cosmological constant and Einstein-Hilbert terms, which correspond to ${\cal L}_2^{\rm (AdS)}$ and ${\cal L}_4^{\rm (AdS)}$ respectively (after using \eqref{gaussdozadse2} to eliminate intrinsic curvatures), and the non-trivial Lovelock invariants in the five dimensional bulk that require boundary terms are the Einstein-Hilbert term and the Gauss-Bonnet term, whose boundary terms are ${\cal L}_3^{\rm (AdS)}$ and ${\cal L}_5^{\rm (AdS)}$ respectively (the cosmological constant term does not require a boundary term).  The term ${\cal L}_1^{\rm (AdS)}$, on the other hand, is the Wess-Zumino term: it cannot be written in terms of the invariant building blocks as in \eqref{dbigbekrlee}, but is still invariant up to a total derivative under the broken conformal symmetries.    It can be written as the 5-volume enclosed by the 3-brane and an arbitrary reference surface \cite{Goon:2011qf}.  Due to the fact that this term starts at linear order in the field and has no derivatives, we call it the tadpole term.

In the Weyl realization, the five galileon terms in $D=4$ are
\bea {\cal L}_1^{\rm (Weyl)} &=& - {1\over 4L^4}  \sqrt{-g}\, ,   \nn\\
 {\cal L}_2^{\rm (Weyl)}  &=&  -{1\over 12L^2 } \sqrt{-g}R \,  ,   \nn\\
 {\cal L}_3^{\rm (Weyl)}   &=&{1\over 2}  (\partial \pi)^2\square\pi +{1\over 4}(\partial\pi)^4 \,, \nn\\
{\cal L}_4^{\rm (Weyl)} &=& L^2   \sqrt{-g}\, {1\over 8}\left( {7\over 36} R^3-R R_{\mu\nu}^2+R_{\mu\nu}^3 \right) \,,\nn\\ 
{\cal L}_5^{\rm (Weyl)}   &=& L^4   \sqrt{-g}\, {1\over 192}\left(  {31\over 18}R^4-13 R_{\mu\nu}^2R^2+9\left(R_{\mu\nu}^2\right)^2+20RR_{\mu\nu}^3-18R_{\mu\nu}^4  \right) \,,\nn\\ 
 \label{weyllagserep2}
\eea
where indices are moved with $g_{\mu\nu}$ and we have used the length scale $L$ of the AdS realization to absorb overall dimensions (which aids when mapping between the realizations).
They are all of the form \eqref{dilatongenformee} except for ${\cal L}_3^{\rm (Weyl)} $: this is the Wess-Zumino term, and it cannot be built in terms of the invariants (the term $\sim R^2$ that gives second order equations of motion ends up vanishing \cite{Nicolis:2008in}).  This term must be built in a different way (e.g. dimensional continuation \cite{Nicolis:2008in} or building an invariant 5-form via the coset construction \cite{Goon:2012dy}).

In \cite{Creminelli:2013fxa}, it was shown that the AdS/CFT equivalence transformation of \cite{Bellucci:2002ji}, when applied to the galileons in $D=4$, relates the Lagrangians \eqref{ghostfreegentermsadsp1} and \eqref{weyllagserep2} via an invertible linear transformation.  This can be expected on the grounds that the galileons are the only terms that non-linearly propagate only a single degree of freedom, and invertible transformations, such as the AdS/CFT equivalence transformation, should preserve the degree of freedom count.

In particular, the AdS/CFT equivalence transformation transforms invariant building blocks on one side into invariant building blocks on the other, so invariant terms of the form \eqref{dbigbekrlee}, \eqref{dilatongenformee}, such as the non-Wess-Zumino galileon terms, are mapped into each other.  The Wess-Zumino terms are also mapped to each other, but they are mapped modulo the addition of invariant terms (this is because the meaning of Wess-Zumino term, i.e. a Lagrangian term not constructed from invariant building blocks and not equivalent to one up to total derivative, but still invariant under the symmetries up to a total derivative, is only well-defined modulo invariant terms).  Thus the fact that there is one Wess-Zumino term among the galileons is independent of the realization of the EFT.

Below we will extend the galileon AdS/CFT equivalence transformation map of \cite{Creminelli:2013fxa} to arbitrary spacetime dimension $D$.  When we do this, we encounter a small puzzle with the Wess-Zumino terms.  In $D$ dimensions there are $D+1$ galileon terms.  In the AdS realization, in arbitrary $D$ all the ${\cal L}_n^{\rm (AdS)} $ for $n>1$ come from various higher Lovelock invariants and boundary terms.  The tadpole term ${\cal L}_1^{\rm (AdS)}$, on the other hand, always has the same structure as in \eqref{weyllagserep2}, ${\cal L}_1^{\rm (AdS)}\sim  e^{D\pi}$ in $D$ dimensions and it can always be written as the $(D+1)$-volume enclosed by the $(D-1)$-brane and an arbitrary reference surface.  In the Weyl realization, for even $D$, there is always a single Wess-Zumino term which is the middle galileon ${\cal L}_{{D\over 2}+1}^{\rm (Weyl)}$, whose would-be ghost free $\sim R^{D/2}$ structure vanishes.  We can expect that this term matches with the tadpole term in the AdS realization (up to non-Wess-Zumino terms).  In odd $D$, on the other hand, there is no middle galileon and no vanishing structures to worry about and all $D+1$ galileons can be constructed as invariant terms.  There is thus no Wess-Zumino term for odd $D$ in the Weyl realization.\footnote{When the breaking of conformal symmetry occurs due to a renormalization group flow, the Wess-Zumino term in the Weyl effective action captures the change in the $a$-type conformal anomaly between the UV and IR fixed points \cite{Komargodski:2011vj,Komargodski:2011xv,Elvang:2012st,Elvang:2012yc}.   The absence of the Wess-Zumino term in odd $D$ reflects the absence of conformal anomalies in odd $D$.}  This raises the puzzle of what happens with the tadpole term ${\cal L}_1^{\rm (AdS)}$ in the AdS realization in odd $D$ and what it corresponds to in the Weyl realization.  
 
 \section{AdS and Weyl galileons in general $D$}
 
 In the Weyl realization, the natural building block that emerges from the coset construction is the quantity $D_\mu\xi_\nu$ defined by \cite{Bellucci:2002ji,McArthur:2010zm,Goon:2012dy,Hinterbichler:2012mv}
 \be
\mathcal D_\mu\xi_\nu =\frac{1}{2}\partial_\mu\pi\partial_\nu\pi-\frac{1}{2}\partial_\mu\partial_\nu\pi-\frac{1}{4}(\partial \pi)^2\eta_{\mu\nu}~.
\label{dxia2}
\ee
Using this is equivalent to using the curvature \eqref{riccite}, the two are related via\footnote{Note that in $D=2$ the relation between $R_{\mu\nu}$ and ${\cal D}_\mu\xi_\nu$ is not invertible.  For this case we must use the more fundamental ${\cal D}_\mu\xi_\nu$ to construct all the galileons: the third galileon is constructed from $S_2[{\cal D}\xi]$, but the only independent curvature invariant at this order is $R^2$, which does not give a galileon.} \be R_{\mu\nu}=2(D-2)\mathcal D_\mu\xi_\nu+2\mathcal D^\rho\xi_\rho g_{\mu\nu}\, ,\ \  D_\mu\xi_\nu={1\over 2(D-2)}\left(R_{\mu\nu}-{1\over 2(D-1)}Rg_{\mu\nu}\right)\, .\ee  
The advantage of using \eqref{dxia2} is that, as is manifest from the coset construction in \cite{Goon:2012dy}, the galileon terms are all symmetric polynomials\footnote{For a $D\times D$ matrix with components $M^\mu_{\ \nu}$, 
the symmetric polynomials for $1\leq n\leq D$ are defined as 
\be S_n[M]=M^{ [\mu_1}_{\ \ \mu_1}\cdots M^{ \mu_n]}_{\ \ \mu_n}\, \ ,\label{symmpoldefeee}\ee
and we take $S_0[M]=1$.  Here the anti-symmetrization is with weight 1. Some properties that will be useful are the shifting identity
\be
S_n(1+M)= \sum_{m=0}^{n}  \binom{D-m}{n-m} S_m(M)\,,
\label{Mshiftingtheorem}
\ee
and if M is invertible, the determinant identity
\be 
S_n(M^{-1})= \frac{S_{D-n}(M)}{\det(M)}\,.
\label{Minversetheorem}
\ee} of the matrix $({\cal D}\xi)^\mu_{\ \nu}=g^{\mu\rho} {\cal D}_\rho\xi_{\nu}$.  In odd $D$, the galileons can be taken as
\bea   
\bar{\cal L}_1^{\rm (Weyl)} &=&L^{-D}   \sqrt{-g}\,S_0[{\cal D}\xi]   \,,\nn\\
\bar{\cal L}_2^{\rm (Weyl)} &=&L^{-D+2}   \sqrt{-g}\,S_1[{\cal D}\xi]   \,,\nn\\
&\vdots & \nn\\
\bar{\cal L}_{D+1}^{\rm (Weyl)} &=& L^D   \sqrt{-g}\,S_D[{\cal D}\xi]   \,,\ \ \ \ \ \ \ \ ({\rm odd}\ D)\,. 
\eea
These are all of the form \eqref{dilatongenformee} and there are no Wess-Zumino terms.

In even $D$, there are an odd number of galileons and the middle galileon ${\cal L}_{{D\over 2}+1}^{\rm (Weyl)} $ is a Wess-Zumino term.  The would-be symmetric polynomial, $S_{D/2}[{\cal D}\xi]$, yields a term that vanishes identically up to a total derivative.  One can construct this term from a $(D+1)$-form in the coset construction\footnote{In the language of \cite{Goon:2012dy}, the Wess-Zumino $(D+1)$-form is $\sim \epsilon_{\mu_1\cdots\mu_D}D\wedge P^{\mu_1}\wedge\cdots\wedge P^{\mu_{D\over 2}}\wedge K^{\mu_{{D\over 2}+1}}\wedge\cdots\wedge K^{\mu_D}$.}, but we will use a different option, the trick used in \cite{Nicolis:2008in}, which is to evaluate $S_{D/2}[{\cal D}\xi]$ in some different spacetime dimension $D'$, whereupon one finds a result proportional to $(D'-D)$.  By stripping off this vanishing factor, one finds a finite result which is the Wess-Zumino term\footnote{See \cite{Bonifacio:2020vbk,Gabadadze:2023quw,Gabadadze:2025uxk} for a connection between this method and the conformal anomaly effective action.},
\be \bar{\cal L}_{{D\over 2}+1}^{\rm (Weyl)} = \lim_{D'\rightarrow D}{1\over D'-D}\left.\left(\sqrt{-g} S_{D/2}[{\cal D}\xi]\right)\right|_{D'\ {\rm dimensions}}\,.\label{WZdefinitnere}\ee
For even $D$, we can therefore write the set of galileon terms as\footnote{Note that in $D=4$ these have a different overall normalization than \eqref{weyllagserep2}, which is why we use the notation $\bar{\cal L}$ rather than ${\cal L}$.}
\bea   
\bar{\cal L}_1^{\rm (Weyl)} &=& L^{-D}   \sqrt{-g}\,S_0[{\cal D}\xi]   \,,\nn\\
&\vdots & \nn\\
\bar{\cal L}_{{D\over 2}}^{\rm (Weyl)} &=& L^{-2}   \sqrt{-g}\,S_{{D\over 2}-1}[{\cal D}\xi]   \,,\nn\\
\bar{\cal L}_{{D\over 2}+1}^{\rm (Weyl)} &=&  \lim_{D'\rightarrow D}{1\over D'-D}\left.\left(\sqrt{-g} S_{D/2}[{\cal D}\xi]\right)\right|_{D'\ {\rm dimensions}}  \,,\nn\\
\bar{\cal L}_{{D\over 2}+2}^{\rm (Weyl)} &=& L^2   \sqrt{-g}\,S_{{D\over 2}+1}[{\cal D}\xi]   \,,\nn\\
&\vdots & \nn\\
\bar{\cal L}_{D+1}^{\rm (Weyl)} &=& L^D   \sqrt{-g}\,S_D[{\cal D}\xi]   \,,\ \ \ \ \ \ \ \ ({\rm even}\ D). 
\eea

For the AdS representation, the natural building block that emerges from the coset construction is the quantity ${\cal D}_\mu\Lambda_\nu$, defined in terms of the extrinsic curvature \eqref{braneexcuree} and brane metric \eqref{adsmetrfee} by
\be
\mathcal D_\mu\Lambda_\nu ={1\over 2L}\left( K_{\mu\nu}-{1\over L}\hat g_{\mu\nu}\right)\,.
\label{dxia22}
\ee
With this, a basis for all the galileons except the first can be chosen to be simply the symmetric polynomials of the matrix $({\cal D}\Lambda)^\mu_{\ \nu}=\hat g^{\mu\rho} {\cal D}_\rho\Lambda_{\nu}$.  We thus take\footnote{In $D=4$ these have a different overall normalization than \eqref{ghostfreegentermsadsp1}, which is why, as in the Weyl case, we use the notation $\bar{\cal L}$ rather than ${\cal L}$.}
\bea   
\bar{\cal L}_1^{\rm (AdS)} &=&L^{-D}{1\over D} e^{D\phi}   \,,\nn\\
\bar{\cal L}_2^{\rm (AdS)} &=& L^{-D}  \sqrt{-\hat g}\,S_0[{\cal D}\Lambda]   \,,\nn\\
&\vdots & \nn\\
\bar{\cal L}_{D+1}^{\rm (AdS)} &=& L^{D-2}   \sqrt{-\hat g}\,S_{D-1}[{\cal D}\Lambda]   \, ,\ \ \ \ \ \ \ \ ({\rm all}\ D).
\eea

Looking at this, an obvious question is why the last symmetric polynomial $S_{D}[{\cal D}\Lambda]$ does not appear.   In even $D$, there is the Lovelock invariant on the brane with $D/2$ powers of the curvature, which is a total derivative.  Expressing this in terms of the symmetric polynomials using \eqref{gaussdozadse2} and \eqref{dxia22}, we can see that there is a linear combination involving $S_{D}[{\cal D}\Lambda]$ that is a total derivative,
\bea {1\over \sqrt{-\hat g}}{\cal L}_{\rm Lovelock}&=&{D!\over 2^{D/2}}\delta^{[\mu_1}_{\nu_1} \delta^{\mu_2}_{\nu_2}\cdots \delta^{\mu_{D-1}}_{\nu_{D-1}}\delta^{\mu_D]}_{\nu_D} \hat R_{\mu_1\mu_2}^{\ \ \   \ \ \nu_1\nu_2}\hat R_{\mu_3\mu_4}^{\ \ \ \ \ \nu_3\nu_4}\cdots \hat R_{\mu_{D-1}\mu_D}^{\ \ \ \ \ \ \ \nu_{D-1}\nu_D} \nn\\
&=& 2^D \left({D\over{2}}\right)! \sum_{m=0}^{D/2} L^{2m}  (1 + m)_{D/2} S_{{D\over{2}}+m}[{\cal D}\Lambda]\,, 
\eea
where the Pochhammer symbol is $(a)_n\equiv a(a+1)(a+2)\cdots(a+n)$ for $n\geq1$, $(a)_0=1$.
We thus have, up to total derivative, 
\bea 0&=&\sqrt{-\hat g} \sum_{m=0}^{D/2} L^{2m}  (1 + m)_{D/2} S_{{D\over{2}}+m}[{\cal D}\Lambda]  \nn\\
&=& \sqrt{-\hat g} L^D {D!\over (D/2)!} S_{D}[{\cal D}\Lambda] + ({\rm lower \ symmetric\ polynomials})\,, \ \ \ ({\rm even}\ D)\,, \nn\\
\eea
and so we can always eliminate $S_{D}[{\cal D}\Lambda]$ in favor of the lower symmetric polynomials.

For odd $D$, we find the resolution to our puzzle about the seeming mismatch of Wess-Zumino terms: the last symmetric polynomial is no longer part of a total derivative, but instead leaves a residual term which gives $\bar{\cal L}_1^{\rm (AdS)}$.  Up to total derivatives we can express $\bar{\cal L}_1^{\rm (AdS)}$ completely in terms of the symmetric polynomials as follows (we find this using the AdS/CFT equivalence relation transformation rules, see section \ref{galmapsection}),
\be
L^{-D}{1\over D} e^{D\phi}
= \sqrt{-\hat g}\sum_{n=0}^D 
{n!\over D \left(1-{D\over 2}\right)_n} L^{2n - D} S_n[{\cal D}\Lambda]\,,\ \ \ ({\rm odd}\ D)\, .  \label{OddDtadmfge}
\ee
The tadpole term is therefore not a Wess-Zumino term in odd $D$, because it can be written as an exactly invariant term up to total derivatives.  (Note that it gets a contribution from the term $\sim K^{D}$ that is the boundary term of the total derivative $\sim \hat R^{D+1\over 2}$ Lovelock term in the even-dimensional bulk.)  This is consistent with the absence of Wess-Zumino terms in the Weyl representation.  For odd $D$, using \eqref{OddDtadmfge}, we can therefore use a basis with only symmetric polynomials,
 \bea   
\tilde{\cal L}_1^{\rm (AdS)} &=&L^{-D}   \sqrt{-g}\,S_0[{\cal D}\Lambda]     \,,\nn\\
\tilde{\cal L}_2^{\rm (AdS)} &=& L^{-D+2}  \sqrt{-g}\,S_1[{\cal D}\Lambda]   \,,\nn\\
&\vdots & \nn\\
\tilde{\cal L}_{D+1}^{\rm (AdS)} &=& L^D  \sqrt{-g}\,S_{D}[{\cal D}\Lambda]   \,,\ \ \ \ \ \ \ \ ({\rm odd}\ D). 
\eea

 \section{Galileon equivalence map in general $D$\label{galmapsection}}

The AdS/CFT equivalence transformation involves a change of coordinates as well as a change in the field, relating $\pi(y)$ and $\phi(x)$.  It is given by
\be y^\mu=x^\mu +L^2e^{-\phi}\lambda^\mu,\ \ \ e^{\pi}={e^{\phi}\over 1+L^2\lambda^2},\ \ \  	\lambda_\mu\equiv-{ e^{-\phi}\partial_\mu \phi\over 1+ \sqrt{ 1+L^2 e^{-2\phi}(\partial \phi)^2}}\,, \ee
or inversely,
\be x^\mu=y^\mu-L^2{e^{-\pi}   \over  1+{ L^2}\xi^2}\xi^\mu\,,  \ \ \  e^{\phi}= \left(1+{ L^2}\xi^2\right) e^\pi, \ \ \ \  \xi_\mu \equiv  -\frac{1}{2}e^{-\pi}\partial_\mu\pi\,, \label{xtoycrelee2}\ee
where indices in these expressions are moved with $\eta_{\mu\nu}$.
This map transforms the invariant building blocks \eqref{dxia2} and \eqref{dxia22} into each other.
In matrix notation, using as the matrices $({\cal D}\xi)^\mu_{\ \nu}=g^{\mu\rho} {\cal D}\xi_{\rho\nu}$ and $({\cal D}\Lambda)^\mu_{\ \nu}=\hat g^{\mu\rho} {\cal D}\Lambda_{\rho\nu}$, this map becomes
\be {\cal D}\xi={1\over 1+L^2 {\cal D}\Lambda} {\cal D}\Lambda,\ \ \ {\cal D}\Lambda={1\over 1-L^2 {\cal D}\xi} {\cal D}\xi\,. \label{invariantrela2tionadese}\ee
In addition, the measure transforms as
\be \sqrt{-g}d^Dy=  \det \left(1+L^2 {\cal D} \Lambda\right)  \sqrt{-\hat g} d^Dx .\label{measuretraadnsfee2}\ee

Using this, along with the properties \eqref{Mshiftingtheorem}, \eqref{Minversetheorem} of the symmetric polynomials, we can see that the symmetric polynomials are mapped into each other under the AdS/CFT equivalence transformation:  
\bea
&&S_n\left[{L^2\cal D}\xi\right] =S_n\left[{L^2\over 1+L^2 {\cal D}\Lambda} {\cal D}\Lambda\right] = S_n\left[1- {1\over 1+L^2 {\cal D}\Lambda} \right]\nn\\
&&  = \sum_{m=0}^{n} (-1)^m \binom{D-m}{n-m}  S_m\left[{1\over 1+L^2 {\cal D}\Lambda}\right] = \sum_{m=0}^{n} (-1)^m \binom{D-m}{n-m}  {S_{D-m}\left[{1+L^2 {\cal D}\Lambda}\right] \over \det( 1+L^2 {\cal D}\Lambda)} \nn\\
&&  = \sum_{m=0}^{n} \sum_{p=0}^{D-m} (-1)^m \binom{D-m}{n-m}  \binom{D-p}{D-m-p}  {S_p\left[L^2 {\cal D}\Lambda\right]\over \det( 1+L^2 {\cal D}\Lambda)} \nn\\
&&= {1\over  \det( 1+L^2 {\cal D}\Lambda)} \sum_{m=n}^{D}\binom{m}{n} S_{m}\left[L^2 {\cal D}\Lambda\right]\,.
\eea
Given that we have chosen a basis consisting of the symmetric polynomials, and using \eqref{measuretraadnsfee2}, this gives a map between galileon actions that do not involve Wess-Zumino terms,
\be  \int d^Dy \sqrt{-g} S_n\left[{L^2\cal D}\xi\right] =  \int  d^Dx \sqrt{-\hat g} \sum_{m=n}^{D}\binom{m}{n} S_{m}\left[L^2 {\cal D}\Lambda\right] \,.\label{gbeffrendgre}\ee

For odd $D$, this gives the complete map since there are no Wess-Zumino terms and all the galileons on both sides can be expressed in terms of symmetric polynomials.  For even $D$, we also need to extend the map to the Wess-Zumino terms.  The easiest way to do this is to directly transform the tadpole term ${\cal L}_1^{(\rm AdS)}$.  Using \eqref{xtoycrelee2} to directly transform $e^{D\phi}\rightarrow\left(1+{ L^2}\xi^2\right)^D e^{D\pi}$, 
and to compute  ${dx^\mu\over dy^\nu }=\delta^\mu_\nu-L^2\partial_\nu\left({e^{-\pi}   \over  1+{ L^2}\xi^2}\xi^\mu\right)\,$ 
and use it to transform the measure $d^Dx=\det \left( dx^\mu\over dy^\nu\right)  d^Dy$, 
the result is a combination of the symmetric polynomials and the Weyl Wess-Zumino terms,
\be
\bar{\cal L}_1^{\rm (AdS)}=L^{-D}{1\over D} e^{D\phi}
= \sqrt{-g} \sum_{n= 0}^D 
\frac{(-1)^n}{D - 2n} \, L^{2n - D}  S_n[{\cal D}\xi] \,.
\ee
For odd $D$, this works as written (and is what gave us \eqref{OddDtadmfge} after using \eqref{gbeffrendgre}).  For even $D$, the term $n=D/2$ in the sum diverges, the signal of the Wess-Zumino term.  For this term, given our definition \eqref{WZdefinitnere}, we should take 
\be \sqrt{-g} S_{D/2}[{\cal D}\xi]\rightarrow\left( {D-2n}\right) \bar {\cal L}_{{D\over 2}+1}^{(\rm Weyl)}\ee 
before evaluating the sum, after which the result will be finite.  This then completes the mapping for even $D$.

\vspace{-5pt}
\paragraph{Acknowledgments:} We thank Alice Garoffolo for discussions.  KH acknowledges support from DOE award DE-SC0009946.

\bibliographystyle{utphys}
\addcontentsline{toc}{section}{References}
\bibliography{generalDdraft_arxiv}

\end{document}